\documentstyle[]{article}
\newlength{\aivwidth}   
\newlength{\tmpwidth}   

\title{ More on scattering of Chern-Simons vortices }
\author{Jacek Dziarmaga  \\
        Jagellonian University, Institute of Physics, \\
        Reymonta 4, 30-059 Krak\'ow, Poland
        \thanks{e-mail address: ufjacekd@ztc386a.if.uj.edu.pl}}
\date{December 20, 1994; revised January 13, 1995}
\begin{document}
\maketitle
    \begin{abstract}
    I derive a general formalism for finding kinetic terms
of the effective Lagrangian for slowly moving Chern-Simons vortices.
Deformations of fields linear in velocities are taken into account.
{}From the equations they must satisfy I extract the kinetic term in the limit
of coincident vortices. For vortices passing one over the other there is
locally the right-angle scattering. The method is based on analysis
of field equations instead of action functional so it may be useful
also for nonvariational equations in nonrelativistic models of
Condensed Matter Physics.
    \end{abstract}
\vspace*{1.5cm}
TPJU 30/94\\
hep-th 9412180\\
\vspace*{1.5cm}
\newcommand{\be}{\begin{equation}\label}
\newcommand{\ee}{\end{equation}}
\newcommand{\ba}{\begin{eqnarray}\label}
\newcommand{\ea}{\end{eqnarray}}
\newcommand{\pl}{\partial}
\newcommand{\lb}{\lambda}
\newcommand{\kp}{\kappa}
\newcommand{\tr}{\triangle}
\newcommand{\dt}{\delta}
\newcommand{\bt}{\beta}
\newcommand{\al}{\alpha}

\section{ Introduction }

    There is quite a large family of field-theoretical models both
relativistic
\cite{jackiwweinberg,jacleewein,hongkimpac,leese,ruback,prassom,lee,kl} and
nonrelativistic \cite{jackiwpi,dunne,igor,ezawa} with
solitons which posses Bogomol'nyi \cite{bogomolny} limit. In this limit
static field equations in a given topological sector can be reduced
for a minimal energy configuration to first order differential equations.
These equations generically admit static multisoliton solutions
characterised by a finite set of parameters such as positions of
solitons and their internal orientations. The configurations are static -
if we think about solitons as about particles there are no net static
forces between them.

    It was an idea of Manton \cite{manton} that low energy scattering of
monopoles
in the Bogomol'nyi-Prasad-Somerfield model \cite{bogomolny,prassom} can
be modeled
by reduction of the dynamics to a finite-dimensional manifold of
parameters of static multisoliton solutions. The kinetic part of the
Lagrangian of the original theory after integrating out spacial
dependence of the fields with by now time-dependent parameters
yields the kinetic part of effective Lagrangian quadratic in
time derivatives of parameters. A metric on the moduli space can be read
out of it. A fundamental idea in this approach is that configurations
satisfying Bogomol'nyi lower bound are at a bottom of a potential well.
A slow motion of solitons can lead only to small deformations
of fields with respect to static configurations.

    The idea was succesfully applied to scattering of monopoles,
vortices in Abelian Higgs model \cite{ruback,samols,rubshe,ja4}, $CP^{1}$
solitons \cite{ward,leese}.
Recently also extensions of the method to the case of Chern-Simons vortices
both relativistic and nonrelativistic were done
\cite{kimmin,kimlee,ja1,ja2,huachou,ja3}.
However as was first pointed out in \cite{kimlee} in these cases a new
problem arises.
Lagrangians of these models contain terms linear in time derivatives
such as Chern-Simons term and/or Schrodinger action.
By just promoting parameters to the role of collective coordinates
one can reliably calculate only terms in the effective Lagrangian linear
in velocities. To compute kinetic part one has to take into
account small deformations of the fields with respect to static
configurations. It is enough to consider only deformations linear
in velocities. Such deformations can be in principle calculated
from full field equations linearised in deformations and terms
linear in velocities. However also terms linear in accelerations
and third time derivatives arise and as I have discussed in \cite{ja2}
there is no appearent reason why they could be negligible as compared
to velocities. Such an "approximation" can lead to serious inconsistencies.

   In this paper I put the problem on a slightly different footing.
The acceleration terms are not neglected. There are no net static
forces between vortices so their accelerations must be zero for
vanishing velocities. Thus we can assume that acceleration vector
is at least linear in velocities. We can expect this linear term to
be nonzero because there are charge-flux interactions by Lorenz-like
forces. Thus acceleration is not neglected but replaced by
a position-dependent matrix $\omega$ times velocity. The same procedure
can be applied iteratively to the third time derivative. Finally
in a special limit of coinciding vortices such a unique form of
$\omega$-matrix is extracted which admits regular deformations of
fields. A knowledge of $\omega$ is enough to establish the form
of moduli space metrics. A brief comment on how this approach
works in Abelian Higgs model is added.

\section{ The model, zero modes and useful notations}

   We take the Lagrangian of the self-dual Chern-Simons-Higgs model
in the form
\be{10}
L=\frac{\kp}{2}\varepsilon^{\mu\nu\alpha}A_{\mu}\pl_{\nu}A_{\alpha}
 +D_{\mu}\phi^{\star}D^{\mu}\phi-V(\mid\phi\mid) \;\;,
\ee
where
$V(\mid\phi\mid)=\frac{1}{\kp^{2}}\mid\phi\mid^{2}(\mid\phi\mid^{2}-1)^{2}$,
$D_{\mu}\phi=\pl_{\mu}\phi-A_{\mu}\phi$, the signature of the flat
2+1 dimentional metrics is $(+,-,-)$ and the Levi-Civita symbol
is choosen so that $\varepsilon^{012}=-1$. A variation of the action
with respect to $A_{0}$ leads to Gauss' law constraint
\be{20}
  2\phi^{\star}\phi(\pl_{t}\chi - A_{0})=\kp B \;\;,
\ee
where we have introduced $\phi=\mid\phi\mid e^{i\chi}$ and the magnetic
field is $B=-F_{12}$. When one takes into account the Gauss' law one
can effectively rewrite the original Lagrangian in a new form
\be{30}
L=\dot{f}^{2} +\kp B\pl_{t}\chi
 +\frac{\kappa}{2}\varepsilon^{ij}A_{i}\pl_{t}A_{j}
 -[\pl_{i}f\pl_{i}f+f^{2}(\pl_{i}\chi-A_{i})^{2}+\frac{\kp^{2}B^{2}}{4f^{2}}
                                +\frac{1}{\kp^{2}}f^{2}(f^{2}-1)^{2}] \;\;.
\ee
We have just introduced $f=\mid\phi\mid$ and $\varepsilon^{ij}$ such that
$\varepsilon^{12}=1$. The energy density for a static configuration
with a positive topological index is
\be{40}
 \varepsilon=[\pl_{i}f-f\varepsilon_{ik}(\pl_{k}\chi-A_{k})]^{2}
     +\frac{1}{f^{2}}[\frac{\kp B}{2}+\frac{f^{2}(1-f^{2})}{\kp}]^{2} - B \;\;.
\ee
The magnetic flux is quantised as $2\pi n$ with $n$ being the winding
number. Thus energy is bounded from below by this value of magnetic
flux. If $\kp>0$ this Bogomol'nyi lower bound is saturated by static
configurations being solutions to first order differential equations
\be{50}
  \pl_{i}f=f\varepsilon_{ik}(\pl_{k}\chi-A_{k}) \;\;,
\ee\be{60}
  \varepsilon_{ij}\pl_{i}A_{j}=\frac{2}{\kp^{2}}f^{2}(1-f^{2}) \;\;,
\ee\be{65}
  A_{0}=\frac{1-f^{2}}{\kp} \;\;.
\ee
These equations admit static multivortex solutions parametrised
by a set of $2n$ real parameters. In Coulomb gauge one can take
$\chi=\sum_{v=1}^{n} Arg(z-z_{v})$, where the sum runs over
vortices labeled by $v$'s and complex coordinates on the plane are used.
In this gauge there is only one second order differential equation
to solve
\be{70}
  \nabla^{2} \ln f=\frac{2}{\kp^{2}}f^{2}(f^{2}-1)
                  +2\pi \sum_{v=1}^{n} \delta^{(2)}(z-z_{v})  \;\;.
\ee
Once $f$ is known other fields can be obtained from Eqs.(\ref{20},\ref{50}).
The singular sources on the R.H.S. of the above equation enforce modulus $f$
in a close vicinity of p-fold zero $z_{0}$ to behave like
$\mid z-z_{0} \mid^{p}$ (it is a leading term of the expansion).
This equation in particular admits cilindrically symmetric vortex
solution with winding number $n$ $\phi=f(r)\exp in\theta$.
We will parametrise multivortex solutions in following equivalent forms
\be{80}
 \phi=(z-z_{1})...(z-z_{n})W(z,\bar{z},z_{v})\equiv
      (z^{n}-\sum_{k=0}^{n-1}\lambda_{k}z^{k})W(z,\bar{z},z_{v}) \;\;
\ee
and
\be{90}
 \chi=\frac{1}{2i}
     \ln\frac{\prod_{v=1}^{n}(z-z_{v})}{\prod_{w=1}^{n}(\bar{z}-\bar{z}_{w})}
\equiv\frac{1}{2i}
     \ln\frac{z^{n}-\sum_{k=0}^{n-1}\lb_{k}z^{k}}
             {\bar{z}^{n}-\sum_{l=0}^{n-1}\bar{\lb}_{l}\bar{z}^{l}} \;\;.
\ee
$\lb$'s are complex coefficients of the n-th degree polynomial with
roots $z_{v}$ and $W$ is a positive real function. Although the
parametrisations
are equivalent it will later appear that $\lb$'s are more efficient
in description of vortices passing one over another.

  Each of these multivortex solutions posesses $2n$ zero modes.
\ba{100}
  \delta f(z,\lb)&=&\frac{\pl f}{\pl\lb^{A}_{p}}(z,\lb)\delta\lb^{A}_{p}\equiv
                       f(z,\lb)h^{A}_{p}(z,\lb)\delta\lb^{A}_{p}
\nonumber\;\;,\\
  \delta\chi(z,\lb)&=&\frac{\pl\chi}{\pl\lb^{A}_{p}}\delta\lb^{A}_{p}=
     -Im(\frac{l^{A}z^{p}}{z^{n}-\lb_{s}z^{s}})
\delta\lb^{A}_{p}\nonumber\;\;,\\
  \delta A_{k}&=&\pl_{k}\delta\chi+\varepsilon_{kl}\pl_{l}\frac{\delta f}{f}=
      (\pl_{k}\frac{\pl\chi}{\pl\lb^{A}_{p}}+\varepsilon_{kl}\pl_{l}h^{A}_{p})
\delta\lb^{A}_{p}\nonumber\;\;,\\
  \delta A_{0}&=&-\frac{2}{\kp^{2}}f^{2}h^{A}_{p}\delta\lb^{A}_{p}\;\;,
\ea
where $\lb_{p}=\lb^{1}_{p}+i\lb^{2}_{p}$ and $l^{A}=(1,i)$.
Equation (\ref{70}) linearised in fluctuations becomes
\be{110}
  \nabla^{2}h^{A}_{p}+\frac{4}{\kp^{2}}\rho(1-2\rho)h^{A}_{p}=0 \;\;.
\ee
$\rho$ denotes moduli squared $f^{2}$. From the asymptotics of $f$ close
to its zeros we can extract the leading term in fluctuations
\be{120}
  h^{A}_{p}\sim -Re(\frac{l^{A}z^{p}}{z^{n}-\lb_{s}z^{s}})
\ee
as $(z^{n}-\lb_{s}z^{s})\sim 0$. This is the only singular term
in the expansion around the actual zero of the Higgs field. This singularity
is fine-tuned by the singularity in $\delta\chi$ to yield regular
$\delta A_{k}$ (see (\ref{100})).

  Now because of future applications let us take a closer look at
the coincident n-vortex solution $\phi=f(r)\exp in\theta$ with $f(r)$
satisfying
\be{130}
  ff''+\frac{ff'}{r}-(f')^{2}=\frac{2}{\kp^{2}}f^{4}(f^{2}-1)
\ee
with boundary conditions $f(0)=0$ and $f(\infty)=1$.
Close to zero it behaves like
$f\sim f_{0}r^{n} -\frac{f_{0}^{3}}{2\kp^{2}(n+1)^{2}}r^{3n+2}+...$.
Fluctuations for small $\lb$'s can be written as
\be{140}
  h(r,\theta)=h^{A}_{p}(r,\theta)\lb^{A}_{p}=
    -H_{n-p}[\lb^{1}_{n-p}\cos(n-p)\theta + \lb^{2}_{n-p}\sin(n-p)\theta] \;\;.
\ee
$H$'s satisfy following equations
\be{150}
  (\frac{d^{2}}{dr^{2}}+\frac{1}{r}\frac{d}{dr}-\frac{(n-p)^{2}}{r^{2}})H_{n-p}
  +\frac{4}{\kp^{2}}\rho(1-2\rho)H_{n-p}=0 \;\;,
\ee
with a normalisation that close to zero $H_{m}\sim\frac{1}{r^{m}}$
and it asymptotically vanishes at infinity.

\section{ Slowly moving vortices }

  For more clarity in this paragraph and in what follows we will rescale
gauge fields $A_{\mu}\rightarrow \kp^{-1}A_{\mu}$ and coordinates
$x^{\mu}\rightarrow \kp^{-1}x^{\mu}$. On the level of field equations
it amounts to fixing $\kp=+1$.

  The aim of this paper is to investigate slow motion of vortices
in adiabatic approximation. In the case of Chern-Simons solitons
a new difficulty arises because there are terms linear in time
derivatives in the Lagrangian of the theory. Only terms linear in
velocities can be correctly calculated by direct application
of former methods. As was pointed out in \cite{kimlee} to obtain the kinetic
term
one has not only to make the static fields time-dependent by time-variation
of their $2n$ parameters but one also has to take into account deformations
of the "static" fields up to terms linear in velocities. It is something
like a generalised Lorenz transformation. In this paragraph we will
develope general formalism for calculating such corrections. It will
appear that to have regular solutions one can not neglect terms
proportional to accelerations as it was anticipated in \cite{ja2}.

   Let us make the fields time dependent by variation of parameters
and add to them deformations linear in velocities,
say $f(z,\lb)\rightarrow f[z,\lb(t)]+\triangle f[z,\lb(t),\dot{\lb}(t)]$ etc.
Evaluation of the effective Lagrangian goes by substitution of such fields
into Lagrangian (\ref{30}) and integrating out their planar dependence.
The only terms which can contribute to the part of the effective Lagrangian
linear in velocities are
\be{170}
  L^{(1)}_{eff}=\int d^{2}x\;
         [B\dot{\chi}+\frac{1}{2}\varepsilon^{ij}A_{i}\dot{A}_{j}] \;\;.
\ee
and one needs to take into account only effects of promoting parameters
to the role of collective coordinates. All the other terms are quadratic
in velocities. The contribution of the second term in Eq.(\ref{170})
vanishes in Coulomb gauge. For fairly separated vortices one can approximate
under the integral $B=-2\pi\sum_{v}\delta^{(2)}(z-z_{v})$ to obtain
\be{180}
  L^{(1)}_{eff}\approx -2\pi\frac{d}{dt}\sum_{v>w} Arg(z_{v}-z_{w}) \;\;.
\ee
With the explicite form of the phase (\ref{90}) and some integration by parts
one can rewrite the linear Lagrangian as
\be{190}
  L^{(1)}_{eff}=-2\pi\sum_{v}\dot{x}_{v}^{i}A_{i}(z_{v}) \;\;,
\ee
where $z_{v}=x^{1}_{v}+ix^{2}_{v}$.

The second order term of the effective Lagrangian is
\ba{220}
  L_{eff}^{(2)}&=&\dot{f}^{2}+2\dot{f}\dot{\tr f}+\dot{\tr f}\dot{\tr f}
                   -(\pl_{i}\tr f)^{2}-\frac{1}{2}V''(f)(\tr f)^{2} \nonumber\\
      &+& f^{2}(\dot{\chi}-\tr A_{0})^{2} -f^{2}(\tr A_{i})^{2}
        + (\tr f)^{2}A_{0}^{2}-(\tr f)^{2}(\pl_{i}\chi-A_{i})^{2}   \nonumber\\
      &+& \tr A_{0} \tr B
        +\frac{1}{2}\varepsilon_{ij}
        (2\dot{A}_{i}\tr A_{j}+\dot{\tr A_{i}}\tr A_{j}) \nonumber\\
      &-& 4f\tr fA_{0}(\dot{\chi}-\tr A_{0})
        + 4f\tr f(\pl_{i}\chi-A_{i})\tr A_{i} \;\;.
\ea
The small corrections to the fields has to be calculated from following
equations obtained by linearisation of full field equations.
\ba{230}
\nabla^{2}(\frac{\tr f}{f})+4\rho(2-3\rho)(\frac{\tr f}{f})+
(\pl_{k}\ln\rho)
[\pl_{k}(\frac{\tr f}{f})+\varepsilon_{kl}(\tr A_{l}-\pl_{l}\tr\chi)]-
2(\rho-1)(\tr A_{0}-\dot{\chi})&=& \frac{\ddot{f}}{f}+
2(\rho-1)\dot{\tr\chi}+\frac{\ddot{\tr f}}{f}                 \nonumber\;\;\\
\nabla^{2}\tr\chi-\pl_{k}\tr A_{k}-2(2\rho-1)\frac{\dot{f}}{f}+
(\pl_{k}\ln\rho)
(\varepsilon_{kl}\pl_{l}\frac{\tr f}{f}+\pl_{k}\tr\chi-\tr A_{k})&=&
\ddot{\chi}-\dot{\tr A_{0}}+\ddot{\tr\chi}
                        -2(1-\rho)\frac{\dot{\tr f}}{f}       \nonumber\;\;\\
\varepsilon_{kl}\pl_{k}\tr A_{l}-4f(1-\rho)\tr f+2\rho(\dot{\chi}-\tr A_{0})
&=&-2\rho\dot{\tr\chi}                                        \nonumber\;\;\\
\pl_{n}\tr A_{0}+2\pl_{n}\rho \frac{\tr f}{f}
+2\rho\varepsilon_{nk}(\pl_{k}\tr\chi-\tr A_{k})-\pl_{n}\dot{\chi}
-\varepsilon_{nk}\pl_{k}(\frac{\dot{f}}{f})&=& \dot{\tr A_{n}}  \;\;.
\ea
The equations were simplified with a use of static field equations
satisfied by background fields. Now the crucial observation is that in
the Bogomol'nyi limit any forces exerted on vortices must be zero for
vanishing velocities so accelerations are at least linear in velocities
\be{240}
  \ddot{\lb}^{A}_{p}=\omega^{AB}_{pq}(\lb)\dot{\lb}^{B}_{q} \;\;.
\ee
$\omega$ for given $p$ is a matrix in indices $A,B$. This relation
can be iterated to give
\be{250}
  \frac{d^{3}}{dt^{3}}{\lb}^{A}_{p}=\omega^{AB}_{pq}(\lb)\ddot{\lb}^{B}_{q}
    +\frac{\pl\omega^{AB}_{pq}}{\pl\lb^{C}_{r}}
                                \dot{\lb}^{B}_{q}\dot{\lb}^{C}_{r}\approx
            \omega^{AB}_{pq}\omega^{BC}_{qr}\dot{\lb}^{C}_{r}   \;\;,
\ee
where once again we have preserved only terms linear in velocities. We can
also make following definitions and approximations
\ba{260}
  \dot{f}&=&fh^{A}_{p}\dot{\lb}^{A}_{p}
\nonumber\;\;,\\
  \ddot{f}&\approx&fh^{A}_{p}\omega^{AB}_{pq}\dot{\lb}^{B}_{q}
\nonumber\;\;,\\
  \tr f&\equiv&fs^{A}_{p}\dot{\lb}^{A}_{p}
\nonumber\;\;,\\
  \dot{\tr
f}&\approx&fs^{A}_{p}\omega^{AB}_{pq}\dot{\lb}^{B}_{q}\nonumber\;\;,\\
  \ddot{\tr f}&\approx&fs^{A}_{p}\omega^{AC}_{pr}
\omega^{CB}_{rq}\dot{\lb}^{B}_{q}\nonumber\;\;,\\
  \ddot{\chi}&\approx&\frac{\pl\chi}{\pl\lb^{A}_{p}}
\omega^{AB}_{pq}\dot{\lb}^{B}_{q}\nonumber\;\;,\\
  \tr A_{k}&\equiv&A_{k}^{Ap}\dot{\lb}^{A}_{p}
\nonumber\;\;,\\
  \tr A_{0}&\equiv&a^{A}_{p}\dot{\lb}^{A}_{p}
\nonumber\;\;,\\
\ea
With these formulas and with the gauge $\tr\chi=0$ Eqs.(\ref{230}) can be
rewritten as
\ba{270}
\dot{\lb}^{A}_{p}[\nabla^{2}s^{A}_{p}+4\rho(2-3\rho)s^{A}_{p}+
     (\pl_{k}\ln\rho)(\pl_{k}s^{A}_{p}+\varepsilon_{kl}A^{Ap}_{l})-
     2(\rho-1)(a^{A}_{p}-\frac{\pl\chi}{\pl\lb^{A}_{p}})]&=&
h^{A}_{p}\omega^{AB}_{pq}\dot{\lb}^{B}_{q}+
s^{A}_{p}\omega^{AC}_{pr}\omega^{CB}_{rq}\dot{\lb}^{B}_{q}      \nonumber\;\;\\
\dot{\lb}^{A}_{p}[-\pl_{k}A_{k}^{Ap}-2(2\rho-1)h^{A}_{p}+
(\pl_{k}\ln\rho)(\varepsilon_{kl}\pl_{l}s^{A}_{p}-A^{Ap}_{k})]&=&
[\frac{\pl\chi}{\pl\lb^{A}_{p}}-a^{A}_{p}-2(1-\rho)s^{A}_{p}]
                       \omega^{AB}_{pq}\dot{\lb}^{B}_{q}       \nonumber\;\;\\
\dot{\lb}^{A}_{p}[\varepsilon_{kl}\pl_{k}A_{l}^{Ap}-4\rho(1-\rho)s^{A}_{p}+
  2\rho(\frac{\pl\chi}{\pl\lb^{A}_{p}}-a^{A}_{p})]&=&0         \nonumber\;\;\\
\dot{\lb}^{A}_{p}[\pl_{n}a^{A}_{p}+2\pl_{n}\rho s^{A}_{p}
                                             -2\rho\varepsilon_{nk}A_{k}^{Ap})
    -(\pl_{n}\frac{\pl\chi}{\pl\lb^{A}_{p}}+\varepsilon_{nk}\pl_{k}h^{A}_{p})]
&=&A^{Ap}_{n}\omega^{AB}_{pq}\dot{\lb}^{B}_{q}                  \;\;.
\ea
They should yield field deformations in approximation up to terms
linear in velocities. In the limit of very small velocities field
deformations become small as compared to background fields so our
linearisations of field equations with respect to deformations are
justified. Because full field deformations are regular functions
then in the limit of slow motion also Eqs.(\ref{270}) must have
regular solutions being good approximations to full deformations.
With this in mind we can take for granted existence of regular
solutions and derive necessary conditions for the regularity.
The unknown parameters in Eqs.(\ref{270}) are elements
of matrices $\omega^{AB}_{pq}(\lb)$. Our strategy from now on is to adjust
such unique values of these parameters which allow solutions to be regular.
Once we know the parameters we will also know equations of
motion for $\lb$'s linearised in velocities
\be{280}
  \ddot{\lb}^{A}_{p}=\omega^{AB}_{pq}\dot{\lb}^{B}_{q} \;\;.
\ee
Since we know linear part of the effective Lagrangian (\ref{180}) the knowledge
of these equations enables us to restore also its quadratic part up to total
derivatives. The general form of the Lagrangian is
\be{290}
 L_{eff}=g^{AB}_{pq}(\lb)\dot{\lb}^{A}_{p}\dot{\lb}^{B}_{q}+
         b^{AB}_{pq}(\lb)\lb^{A}_{p}\dot{\lb}^{B}_{q}  \;\;.
\ee
The metric tensor on the moduli space $g$ must be symmetric
under exchange of pairs of indices $(A,p)$ and $(B,q)$ and invertible
\be{300}
 (g^{-1})^{AB}_{pq}g^{BC}_{qr}=\dt^{AC}\dt_{pr} \;\;.
\ee
Equations of motion linearised in velocities are
\be{310}
 2g^{AB}_{pq}\ddot{\lb}^{A}_{p}+
 (b^{AB}_{pq}-b^{BA}_{qp})\dot{\lb}^{A}_{p}+
 (\frac{\pl b^{AB}_{pq}}{\pl\lb^{D}_{s}}
 -\frac{\pl b^{AD}_{ps}}{\pl\lb^{B}_{q}})\lb^{A}_{p}\dot{\lb}^{D}_{s} \;\;.
\ee
The components of the metric tensor must solve following equations
\be{320}
g^{BC}_{qr}\omega^{CA}_{rp}+\frac{1}{2}(b^{AB}_{pq}-b^{BA}_{qp})+
\frac{1}{2}(\frac{\pl b^{CB}_{rq}}{\pl\lb^{A}_{p}}
           -\frac{\pl b^{CA}_{rp}}{\pl\lb^{B}_{q}})\lb^{C}_{r}=0.
\ee
This is a system of $2n\times 2n$ linear inhomogenous equations. The
basic condition for the system to have an unique solution is that matrix
$\omega$ is invertible in pairs of indices $(A,p)$ and $(C,r)$,
$det\omega\neq 0$.
This means (see Eq.(\ref{280})) that whatever is the small velocity
it is always a source of acceleration already in linear terms. It should
be a quite generic case except some anomalous sets of measure zero
in a model with magnetic interaction. The metrics can be extended
to these exceptional points by continuity. One of such points is certainly
the limit of infinitely separated vortices where magnetic interaction
degenerates
to a purely topological term (Eq.(\ref{180}). But in this limit the
effective Lagrangian can be accurately calculated with a help
of the product Ansatz of independently Lorenz-boosted vortices.
The quadratic term reads
\be{330}
  L^{(2)}_{eff}\approx\pi\sum_{v}\dot{z}_{v}\dot{z}_{v}^{\star}  \;\;.
\ee
The second and third terms in Eq.(\ref{320}) are explicitely antisymmetric
under exchange of pairs $(A,p)$ and $(B,q)$ so the first term also has to be
antisymmetric $\omega^{T}g=-g\omega$. This condition means that to leading
order acceleration is orthogonal to velocity with respect to metric $g$.
Once again it should be so for forces due to magnetic interactions and they
are the only forces linear in velocities. If this condition is satisfied
we are left with $2n^{2}-n$ independent equations necessary to establish
the same number of metric tensor's components.

\section{ Two vortices in center of mass frame }

  We will consider the by now classic example of two vortices in CM frame. The
system is well described by two parameters $\lb_{1}$ and $\lb_{2}$ which
can be identified with former notations
\be{500}
  \lb_{1}=\lb^{1}_{0} \;\;\;,\;\;\; \lb_{2}=\lb^{2}_{0} \;\;.
\ee
The Higgs field is $\phi=(z^{2}-\lambda)W(z,\bar{z},\lambda)$.
Eqs.(\ref{270}) are on both sides linear in velocities so they can be cast
in a form of a velocity-independent matrix multiplying a vector of velocities.
For the product
to be always zero any velocity vector must belong to kernels of the
matrices. A matrix is zero
if and only if it anihilates any vector from the basis spaning the space
of velocities. If we choose $\dot{\lb}_{1}\neq 0$
and $\dot{\lb}_{2}=0$ Eqs.(\ref{270}) reduce to
\ba{510}
\nabla^{2}s^{(1)}+4\rho(2-3\rho)s^{(1)}+
     (\pl_{k}\ln\rho)(\pl_{k}s^{(1)}+\varepsilon_{kl}A^{(1)}_{l})-
     2(\rho-1)(a^{(1)}-\frac{\pl\chi}{\pl\lb^{(1)}})]&=&
     h^{(A)}\omega^{A1}+s^{(B)}\omega^{BA}\omega^{A1}          \nonumber\;\;\\
-\pl_{k}A_{k}^{(1)}-2(2\rho-1)h^{(1)}+
(\pl_{k}\ln\rho)(\varepsilon_{kl}\pl_{l}s^{(1)}-A^{(1)}_{k})&=&
[\frac{\pl\chi}{\pl\lb^{(A)}}-a^{(A)}-2(1-\rho)s^{(A)}]\omega^{A1}
                                                               \nonumber\;\;\\
\varepsilon_{kl}\pl_{k}A_{l}^{(1)}-4\rho(1-\rho)s^{(1)}+
  2\rho(\frac{\pl\chi}{\pl\lb_{1}}-a^{(1)})&=&0         \nonumber\;\;\\
\pl_{n}a^{(1)}+2\pl_{n}\rho s^{(1)}-2\rho\varepsilon_{nk}A_{k}^{(1)}
    -(\pl_{n}\frac{\pl\chi}{\pl\lb^{(1)}}+\varepsilon_{nk}\pl_{k}h^{(1)})]
&=&A^{(A)}_{n}\omega^{A1}             \;\;.
\ea
On the other hand for the choice $\dot{\lb}_{1}=0$ and
$\dot{\lb}_{2}\neq0$ we obtain
\ba{512}
\nabla^{2}s^{(2)}+4\rho(2-3\rho)s^{(2)}+
     (\pl_{k}\ln\rho)(\pl_{k}s^{(2)}+\varepsilon_{kl}A^{(2)}_{l})-
     2(\rho-1)(a^{(2)}-\frac{\pl\chi}{\pl\lb^{(2)}})]&=&
     h^{(A)}\omega^{A2}+s^{(B)}\omega^{BA}\omega^{A2}          \nonumber\;\;\\
-\pl_{k}A_{k}^{(2)}-2(2\rho-1)h^{(2)}+
(\pl_{k}\ln\rho)(\varepsilon_{kl}\pl_{l}s^{(2)}-A^{(2)}_{k})&=&
[\frac{\pl\chi}{\pl\lb^{(A)}}-a^{(A)}-2(1-\rho)s^{(A)}]\omega^{A2}
                                                               \nonumber\;\;\\
\varepsilon_{kl}\pl_{k}A_{l}^{(2)}-4\rho(1-\rho)s^{(2)}+
  2\rho(\frac{\pl\chi}{\pl\lb_{2}}-a^{(2)})&=&0         \nonumber\;\;\\
\pl_{n}a^{(2)}+2\pl_{n}\rho s^{(2)}-2\rho\varepsilon_{nk}A_{k}^{(2)}
    -(\pl_{n}\frac{\pl\chi}{\pl\lb^{(2)}}+\varepsilon_{nk}\pl_{k}h^{(2)})]
&=&A^{(A)}_{n}\omega^{A2}             \;\;.
\ea
These two sets of equations have to be satisfied simultaneously.

   Now let us consider the limit of coincident vortices. Such a configuration
is rotationally symmetric and it should be all the same what is the
direction in which it is splitting. This motivates the limiting form
of the effective Lagrangian
\be{513}
  L_{eff}^{\lb\rightarrow 0}\sim
    h_{0}\frac{1}{2}\mid\lb\mid^{\xi}\dt^{AB}\dot{\lb}_{A}\dot{\lb}_{B}
                 -g_{0}\varepsilon^{AB}\lb_{A}\dot{\lb}_{B} \;\;,
\ee
with the coefficient $h_{0}$ and the power $\xi$ to be determined.
$g_{0}$ can be extracted with a help of Eq.(\ref{170}) and fluctuations
(\ref{100},\ref{140})
\be{514}
g_{0}=4\pi\int_{0}^{\infty} dr\; \frac{\rho(1-2\rho)H_{2}(r)}{r}
\ee
A value of $g_{0}$ was estimated numerically to be $g_{0}\approx 0.0194$.

   Motivated by null powers of $\lb$'s in Eqs.(\ref{500},\ref{510})
and by
\be{520}
\frac{\pl\chi}{\pl\lb_{1}}=\frac{\sin 2\theta}{r^{2}} \;\;\;,\;\;\;
\frac{\pl\chi}{\pl\lb_{2}}=-\frac{\cos 2\theta}{r^{2}}\;\;
\ee
in the limit of vanishing $\lb$'s we can restrict field deformations
to following forms in polar coordinates $r,\theta$
\ba{530}
s^{(1)}&=&s(r)\sin 2\theta \;\;,\;\;
s^{(2)}=-s(r)\cos 2\theta \nonumber\;\;,\\
a^{(1)}&=&a(r)\sin 2\theta \;\;,\;\;
a^{(2)}=-a(r)\cos 2\theta \nonumber\;\;,\\
A^{(1)}_{r}&=&b(r)\cos 2\theta \;\;,\;\;
A^{(2)}_{r}=b(r)\sin 2\theta \nonumber\;\;,\\
A^{(1)}_{\theta}&=&c(r)\sin 2\theta \;\;,\;\;
A^{(2)}_{\theta}=-c(r)\cos 2\theta \;\;.
\ea
together with a form of the matrix
$\omega^{AB}=\omega\varepsilon^{AB}$. Eqs.(\ref{510}) and (\ref{512})
reduce to
\ba{540}
s''+\frac{s'}{r}-\frac{4s}{r^{2}}+\omega^{2}s+4\rho(2-3\rho)s+
\frac{\rho'}{\rho}(s'+c)-2(\rho-1)(a-\frac{1}{r^{2}})
                                           &=&\omega H_{2}  \nonumber\;\;,\\
b'+\frac{b}{r}+\frac{2c}{r}-2(2\rho-1)H_{2}
+\frac{\rho'}{\rho}(b-\frac{2s}{r})&=&
-\frac{\omega}{r^{2}}+\omega a-2\omega(\rho-1)s             \nonumber\;\;,\\
c'+\frac{c}{r}+\frac{2b}{r}-4\rho(1-\rho)s+
2\rho(\frac{1}{r^{2}}-a)&=&0                                \nonumber\;\;,\\
a'+2\rho's-2\rho c+\omega b&=&
                           -\frac{2}{r^{3}}+\frac{2H_{2}}{r}\nonumber\;\;,\\
\frac{2a}{r}+2\rho b-\omega c&=&\frac{2}{r^{3}}+H_{2}'                \;\;.
\ea
$\omega$ is an adjustable parameter we have to choose in such a way that
solutions are regular. Short inspection shows that fourth equation
in the above set of equations can be derived from the fifth, second and
third. Regularity of $\tr f$ means that $s$ can not be
more divergent then $O(r^{-2})$. Thus we can expand the regular solution
around $r=0$ as
\ba{550}
s(r)=\sum_{k=-2}^{\infty}s_{k}r^{k} \;\;,\;\;
a(r)=\sum_{k=0}^{\infty}a_{k}r^{k} \;\;,\;\;
b(r)=\sum_{k=1}^{\infty}b_{k}r^{k} \;\;,\;\;
c(r)=\sum_{k=1}^{\infty}c_{k}r^{k} \;\;.
\ea
Substitution of first few terms in the expansions to Eqs.(\ref{540})
shows that they are in contradiction unless $\omega=-2$. If we adopt
this value of $\omega$ following leading terms will be obtained
\ba{560}
s&=& (-v-\xi) r^{2} +... \nonumber\;,\\
a&=& (-v-2\xi) r^{2} +...  \nonumber\;,\\
b&=& (-2v-2\xi) r +...       \nonumber\;,\\
c&=& (2v+2\xi) r       \;\;,
\ea
where $v\approx 1387000$ is a coefficient in the expansion
$H_{2}=r^{-2}+vr^{2}+...\;$ and $\xi$ is a free parameter coming from
a "homogenous" part of the solution. For the solution regular at infinity
numerical analysis has given $\xi\approx -1387017.5$.

  With the value of $\omega=-2$ we can conclude that the limiting
form of the effective Lagrangian for $\lambda\rightarrow 0$ is
\be{570}
  L_{eff}^{\lb\rightarrow 0}=
    g_{0}(\frac{1}{2}\dt^{AB}\dot{\lb}_{A}\dot{\lb}_{B}
                               -\varepsilon^{AB}\lb_{A}\dot{\lb}_{B}) \;\;.
\ee

In a generic case of two vortices in CM frame at $(R,\Theta)$ and
$(R,\Theta+\pi)$ the effective Lagrangian must take a form
\be{580}
 L_{eff}=F(R)\dot{R}^{2}+2G(R)\dot{R}R\dot{\Theta}+H(R)R^{2}\dot{\Theta}^{2}
                                                  +B(R)R\dot{\Theta} \;\;.
\ee
The functions $F,G,H$ are to be determined. It is a general form
of the metrics tensor invariant with respect to rotations. The function
$G$ is in general nonzero since the Chern-Simons term breaks parity
invariance. Eq.(\ref{220}) also contains terms which break parity
and there does not seem to be any reason why these terms should vanish.
In polar coordinates the $\omega$-matrix reads
\be{590}
 \omega^{AB}
          =\frac{\frac{d}{dR}[RB(R)]}{2R}(g^{-1})^{AC}\varepsilon^{CB}
          =\frac{J'(R)}{2R}(g^{-1})^{AC}\varepsilon^{CB} \;\;
\ee
with $J(R)$ being the total spin of two vortices separated by
a distance $2R$.
It is an invertible matrix and acceleration is indeed to leading order
orthogonal to velocity. I have attempted calculating
$\omega$ but because there is less symmetry (less constraints) in the problem
for generic $R$ then for $R=0$ it can not be extracted just from the
asymptotices close to zeros of the Higgs field. Matching with asymtotices
at infinity would be necessary.

  Finally a comment on analogous scattering of vortices in Abelian
Higgs model (AHM) is in order. In this model $G(R)=0$ and also $B(R)=0$.
Similar considerations as just above lead to a conclusion that $\omega^{AB}=0$.
Thus in AHM accelerations are at least quadratic in velocities.
In equations analogous to (\ref{230}) terms linear in acceleration
and its time derivative can be neglected as compared to those linear
in velocities. These terms were indeed neglected in Appendix B of
Ref.\cite{kimlee}
and it was shown that rearrangement of effective Lagrangian
analogous to (\ref{220}) with a help of equations fulfilled by deformation
leads to the same expression as that derived by Samols \cite{samols}.
Here we have shown justification of these steps when performed
in AHM.

  Let us consider a general Bogomol'nyi theory and try to decide what
are the conditions for the effective Lagrangian to be purely quadratic
in time derivatives. A Lagrangian of a theory can be written as
\be{600}
  L=G_{ab}[\psi]\dot{\psi_{a}}\dot{\psi_{b}}
               +K_{a}[\psi]\dot{\psi_{a}} -\varepsilon[\psi]
\ee
where $\psi$'s are a set of fields, $\varepsilon[\psi]$ is a static
energy density functional and $G_{ab}[\psi]$ is an invertible, symmetric
and positively definite tensor. The only contribution to the linear part
of the effective Lagrangian is
\be{602}
  L^{(1)}_{eff}=K_{a}[\psi]\dot{\psi_{a}} \;\;,
\ee
where like in all of this paper $\psi$'s are the fields of the static
self-dual background and the time derivative means a total derivative
with respect to time-dependent parameters. This term certainly vanishes if
$K_{a}[\psi]=0$ for the given background. More subtle possibility
is that the whole expression (\ref{602}) can yield zero result when
its spatial dependence is integrated out. Relativistic gauge theory
contains linear terms like in Eq.(\ref{600}). In the Abelian Higgs model
such a term is like
$-\pl_{i}A_{0}\dot{A_{i}}+e\psi^{\star}\psi A_{0}\dot{\chi}$ but the self-dual
background has the property that $A_{0}=0$ and that is why there is no
linear term in the effective Lagrangian. An interesting example of
Maxwell-Higgs self-dual model with relativistic kinetics can be found in
\cite{kl}. A uniform background charge density forces
nonzero $A_{0}$. Vortices in this model feel both Magnus force and mutual
magnetic interactions.

 The quadratic term of the general effective Lagrangian is
\be{610}
  L^{(2)}_{eff}=G_{ab}[\psi](\dot{\psi_{a}}\dot{\psi_{b}} +
  2\dot{\psi_{a}}\dot{\tr\psi_{b}}+\dot{\tr\psi_{a}}\dot{\tr{\psi_{b}}})
  +\frac{\dt K_{a}}{\dt\psi_{b}}
  (\dot{\psi_{a}}\tr\psi_{b}-\tr\psi_{a}\dot{\psi_{b}}
  -\tr{\psi_{a}}\dot{\tr\psi_{b}})
  -\frac{\dt^{2}\varepsilon}{\dt\psi_{a}\dt\psi_{b}}[\psi]
                                   \tr\psi_{a}\tr\psi_{b} \;\;,
\ee
where we have introduced time derivatives of background fields and
deformations of fields. Now we restrict to the case of $L^{(1)}_{eff}=0$.
This means that accelerations are at least quadratic in velocities
and the above formula can be reduced to
\be{611}
  L^{(2)}_{eff}=G_{ab}[\psi]\dot{\psi_{a}}\dot{\psi_{b}}
  +\frac{\dt K_{a}}{\dt\psi_{b}}
  (\dot{\psi_{a}}\tr\psi_{b}-\tr\psi_{a}\dot{\psi_{b}})
  -\frac{\dt^{2}\varepsilon}{\dt\psi_{a}\dt\psi_{b}}[\psi]
                                   \tr\psi_{a}\tr\psi_{b} \;\;,
\ee
We can linearise field equations with respect to
velocities. The field deformations must satisfy a simple equation
\be{613}
  -\frac{\dt^{2}\varepsilon}{\dt\psi_{a}\dt\psi_{b}}\tr\psi_{b}=
  (\frac{\dt K_{a}}{\dt\psi_{b}}-\frac{\dt K_{b}}{\dt\psi_{a}})
                                                    \dot{\psi_{b}}    \;\;.
\ee
This relation enables us to simplify Eq.(\ref{611}) to the following
compact form
\be{620}
  L_{eff}=G_{ab}[\psi]\dot{\psi_{a}}\dot{\psi_{b}}
\ee
plus higher order terms negligible in adiabatic approximation.

   Thus whenever there is no linear term in the effective Lagrangian
the quadratic term can be correctly calculated just with a help
of background fields with their parameters promoted to the role
of collective coordinates.

\section{ Conclusions }

     The limiting form of the term quadratic in velocities of the effective
Lagrangian for two Chern-Simons vortices was extracted from equations
satisfied by deformations of the fields with respect to static background.
This form shows that as we trace locally trajectories of vortices passing
one over another there is the celebrated right-angle scattering.
Globaly if two vortices were pushed from a large distance
one against the other with zero impact parameter they would avoid direct
collision their trajectories being curved by charge-flux interactions.
A difference between a total spin of the pair of vortices when they are
infinitely separated and when they sit on top of one another is $-2\pi$
(for $\kp=1$). Thus the necessary condition for the zeros to meet and
the right-angle scattering to occur is that an impact parameter $d$
with respect to the center of mass and an initial velocity $v$ satisfy
$dv=\frac{1}{2}$. This condition can become sufficient only for
$d$ small enough because vortex-vortex magnetic interactions are
falling exponentially with a distance.

      The local right angle scattering is a hint that moduli-space manifold
is similar to a smoothed cone. The missing volume can show itself
in a thermodynamics of a vortex gas by an excluded volume
in van der Waals state equation similarly as for the vortex gas in
Abelian Higgs model \cite{therm}.

    Finally let me stress that problems partially overcome
in this paper are not at all special to relativistic Chern-Simons models.
They
can also appear in nonrelativistic and nonvariational models so celebrated
in Condensed Matter Physics because Schrodinger or diffusion terms are
linear in time derivatives on one hand and on the other hand the effective
action method may be not able to describe the whole variety of dynamical
phenomena. I would like to address such problems in a near future.

$Acknowledgement.$ I would like to thank Prof. Henryk Arod\'z and
my collegues from Department of Field Theory for discussions during
the preparation of this work. Special thanks should be directed
to Prof. Ki-Myeong Lee for his critical remarks on the first version
of the preprint. This research was supported in part by KBN grant.

\end{document}